\documentclass{PoS}
\usepackage{epsfig} 
\usepackage{wrapfig}

\title{Onset of deconfinement and search for the critical point of strongly interacting matter at CERN SPS energies  }

\ShortTitle{Onset of deconfinement and search for the critical point}

\author{\speaker{Maciej Rybczy\'{n}ski (for the NA49 Collaboration)}\\
        Institute of Physics, Jan Kochanowski University, PL-25406 Kielce, Poland\\
        E-mail: \email{maciej.rybczynski@ujk.edu.pl}}

\author{The NA49 Collaboration:\\
T.~Anti\'ci\'c$^{22}$, B.~Baatar$^{8}$, D.~Barna$^{4}$, J.~Bartke$^{6}$,
H.~Beck$^{9}$, L.~Betev$^{10}$, H.~Bia{\l}\-kowska$^{19}$, C.~Blume$^{9}$,
M.~Bogusz$^{21}$, B.~Boimska$^{19}$, J.~Book$^{9}$, M.~Botje$^{1}$,
P.~Bun\v{c}i\'{c}$^{10}$,
T.~Cetner$^{21}$, P.~Christakoglou$^{1}$,
P.~Chung$^{18}$, O.~Chv\'{a}la$^{14}$, J.G.~Cramer$^{15}$, V.~Eckardt$^{13}$,
Z.~Fodor$^{4}$, P.~Foka$^{7}$, V.~Friese$^{7}$,
M.~Ga\'zdzicki$^{9,11}$, K.~Grebieszkow$^{21}$, C.~H\"{o}hne$^{7}$,
K.~Kadija$^{22}$, A.~Karev$^{10}$, V.I.~Kolesnikov$^{8}$, M.~Kowalski$^{6}$,
D.~Kresan$^{7}$,
A.~L\'{a}szl\'{o}$^{4}$, R.~Lacey$^{18}$, M.~van~Leeuwen$^{1}$,
M.~Ma\'{c}kowiak-Paw{\l}owska$^{9,21}$, M.~Makariev$^{17}$, A.I.~Malakhov$^{8}$,
M.~Mateev$^{16}$, G.L.~Melkumov$^{8}$, M.~Mitrovski$^{9}$, St.~Mr\'owczy\'nski$^{11}$,
V.~Nicolic$^{22}$, G.~P\'{a}lla$^{4}$, A.D.~Panagiotou$^{2}$, W.~Peryt$^{21}$,
J.~Pluta$^{21}$, D.~Prindle$^{15}$,
F.~P\"{u}hlhofer$^{12}$, R.~Renfordt$^{9}$, C.~Roland$^{5}$, G.~Roland$^{5}$,
M. Rybczy\'nski$^{11}$, A.~Rybicki$^{6}$, A.~Sandoval$^{7}$,
N.~Schmitz$^{13}$, T.~Schuster$^{9}$, P.~Seyboth$^{13}$, F.~Sikl\'{e}r$^{4}$,
E.~Skrzypczak$^{20}$, M.~S{\l}odkowski$^{21}$, G.~Stefanek$^{11}$, R.~Stock$^{9}$,
H.~Str\"{o}bele$^{9}$, T.~\v{S}u\v{s}a$^{22}$, M.~Szuba$^{21}$,
M.~Utvi\'{c}$^{9}$, D.~Varga$^{3}$, M.~Vassiliou$^{2}$,
G.I.~Veres$^{4}$, G.~Vesztergombi$^{4}$, D.~Vrani\'{c}$^{7}$,
Z.~W{\l}odarczyk$^{11}$, A.~Wojtaszek-Szwarc$^{11}$}

\author{\\
\vspace{1.0cm}\\
$^{1}$ NIKHEF, Amsterdam, Netherlands. \\
$^{2}$ Department of Physics, University of Athens, Athens, Greece.\\
$^{3}$ E\"otv\"os Lor\'and University, Budapest, Hungary \\
$^{4}$ Wigner Research Center for Physics, Hungarian Academy of Sciences, Budapest, Hungary.\\
$^{5}$ MIT, Cambridge, USA.\\
$^{6}$ H.~Niewodnicza\'nski Institute of Nuclear Physics, Polish Academy of Sciences, Cracow, Poland.\\
$^{7}$ GSI Helmholtzzentrum f\"{u}r Schwerionenforschung GmbH, Darmstadt, Germany.\\
$^{8}$ Joint Institute for Nuclear Research, Dubna, Russia.\\
$^{9}$ Fachbereich Physik der Universit\"{a}t, Frankfurt, Germany.\\
$^{10}$ CERN, Geneva, Switzerland.\\
$^{11}$ Institute of Physics, Jan Kochanowski University, Kielce, Poland.\\
$^{12}$ Fachbereich Physik der Universit\"{a}t, Marburg, Germany.\\
$^{13}$ Max-Planck-Institut f\"{u}r Physik, Munich, Germany.\\
$^{14}$ Institute of Particle and Nuclear Physics, Charles University, Prague, Czech Republic.\\
$^{15}$ Nuclear Physics Laboratory, University of Washington, Seattle, WA, USA.\\
$^{16}$ Atomic Physics Department, Sofia University St. Kliment Ohridski, Sofia, Bulgaria.\\
$^{17}$ Institute for Nuclear Research and Nuclear Energy, BAS, Sofia, Bulgaria.\\
$^{18}$ Department of Chemistry, Stony Brook University (SUNYSB), Stony Brook, USA.\\
$^{19}$ Institute for Nuclear Studies, Warsaw, Poland.\\
$^{20}$ Institute for Experimental Physics, University of Warsaw, Warsaw, Poland.\\
$^{21}$ Faculty of Physics, Warsaw University of Technology, Warsaw, Poland.\\
$^{22}$ Ru{\dj}er Bo\v{s}kovi\'c Institute, Zagreb, Croatia.\\ }

\abstract{
The exploration of the QCD phase diagram particularly the search for a phase transition from hadronic to partonic degrees of freedom and possibly a critical endpoint, is one of the most challenging tasks in present heavy-ion physics. As observed by the NA49 experiment, several hadronic observables in central Pb+Pb collisions at the CERN SPS show qualitative changes in their energy dependence. These features are not observed in elementary interactions and indicate the onset of a phase transition in the SPS energy range. The existence of a critical point is expected to result in the increase of event-by-event fluctuations of various hadronic observables provided that the freeze-out of the measured hadrons occurs close to its location in the phase diagram and the evolution of the final hadron phase does not erase the fluctuations signals. A selection of NA49 results on di-pion and proton intermittency from the scan of the phase diagram will be discussed. 
}

\FullConference{36th International Conference on High Energy Physics,\\
		July 4-11, 2012\\
		Melbourne, Australia}

\begin{document}

\section{Introduction}
In 1999, the NA49 experiment at the CERN Super Proton Synchrotron started a search for the onset of quark-gluon plasma (QGP)~\cite{qgp} creation with data taking for central Pb+Pb collisions
at 40A GeV. Runs at 80A and 20A, 30A GeV followed in 2000 and 2002, respectively. This
search was motivated by the predictions of a statistical model of the early stage of nucleus-nucleus collisions (SMES)~\cite{Gazdzicki:1998vd} that the onset of deconfinement should lead to rapid changes of the energy dependence of numerous hadron production properties, all appearing in a common
energy domain. Conjectured features were observed \cite{Alt:2007aa, Gazdzicki:2010iv} around 30A GeV and dedicated
experiments, NA61/SHINE at the CERN SPS and the beam energy scan at BNL RHIC,
continue detailed studies in the energy region of the onset of deconfinement.

\section{Onset of deconfinement}
A detailed review of the experimental and theoretical status of NA49 evidence for the onset of
deconfinement can be found in the recent review~\cite{Gazdzicki:2010iv}.

Several structures in excitation functions were expected within the SMES:
a kink in the increase of the pion yield per participant nucleon (change of
slope due to increased entropy as a consequence of the activation of partonic
degrees of freedom), a sharp peak (horn) in the strangeness to entropy
ratio, and a step in the inverse slope parameter of transverse mass spectra
(constant temperature and pressure in a mixed phase). Such signatures
were indeed observed in A + A collisions by the NA49 experiment~\cite{Alt:2007aa}, thus
locating the onset of deconfinement energy around 30A GeV ($\sqrt{s_{NN}}\approx 7.6~{\rm GeV}$).

\begin{wrapfigure}{r}{7.0cm}
\includegraphics[scale=.37]{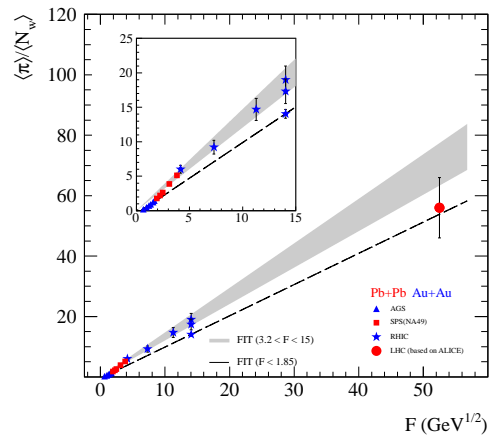}
\caption{\footnotesize Mean pion multiplicity per participant nucleon as function of the
Fermi variable $F \approx s^{0.25}$.}
\label{fig1}
\end{wrapfigure}

\subsection{Verification of NA49 results and interpretation by STAR and ALICE}
Until recently the evidence of onset of deconfinement was based on the results of a single
experiment. Recently new results on central Pb+Pb collisions at the LHC \cite{Schukraft:2011cz} and data on central Au+Au collisions from the RHIC BES program \cite{Kumar:2011us} were released. Figure~\ref{fig1} shows an update of the kink plot, where BES points follow the line for A+A collisions and the LHC point\footnote{The mean pion multiplicity at LHC was estimated based on the ALICE measurement of charged particle multiplicity, see~\cite{footnote} for details.}, within a large error, does not contradict extrapolations from high SPS and RHIC energies.

\begin{figure}
\includegraphics[width=.52\textwidth]{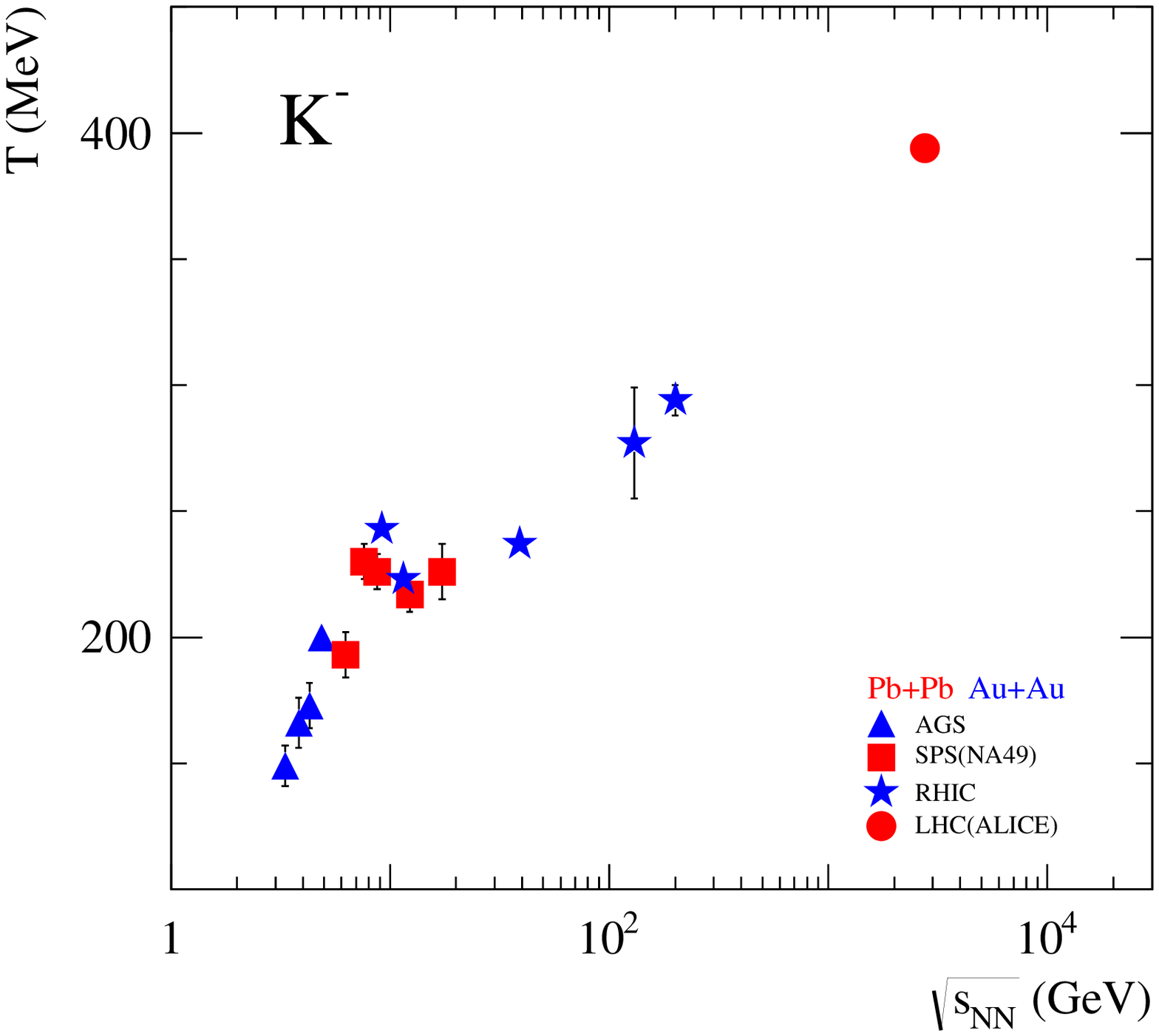}
\includegraphics[width=.52\textwidth]{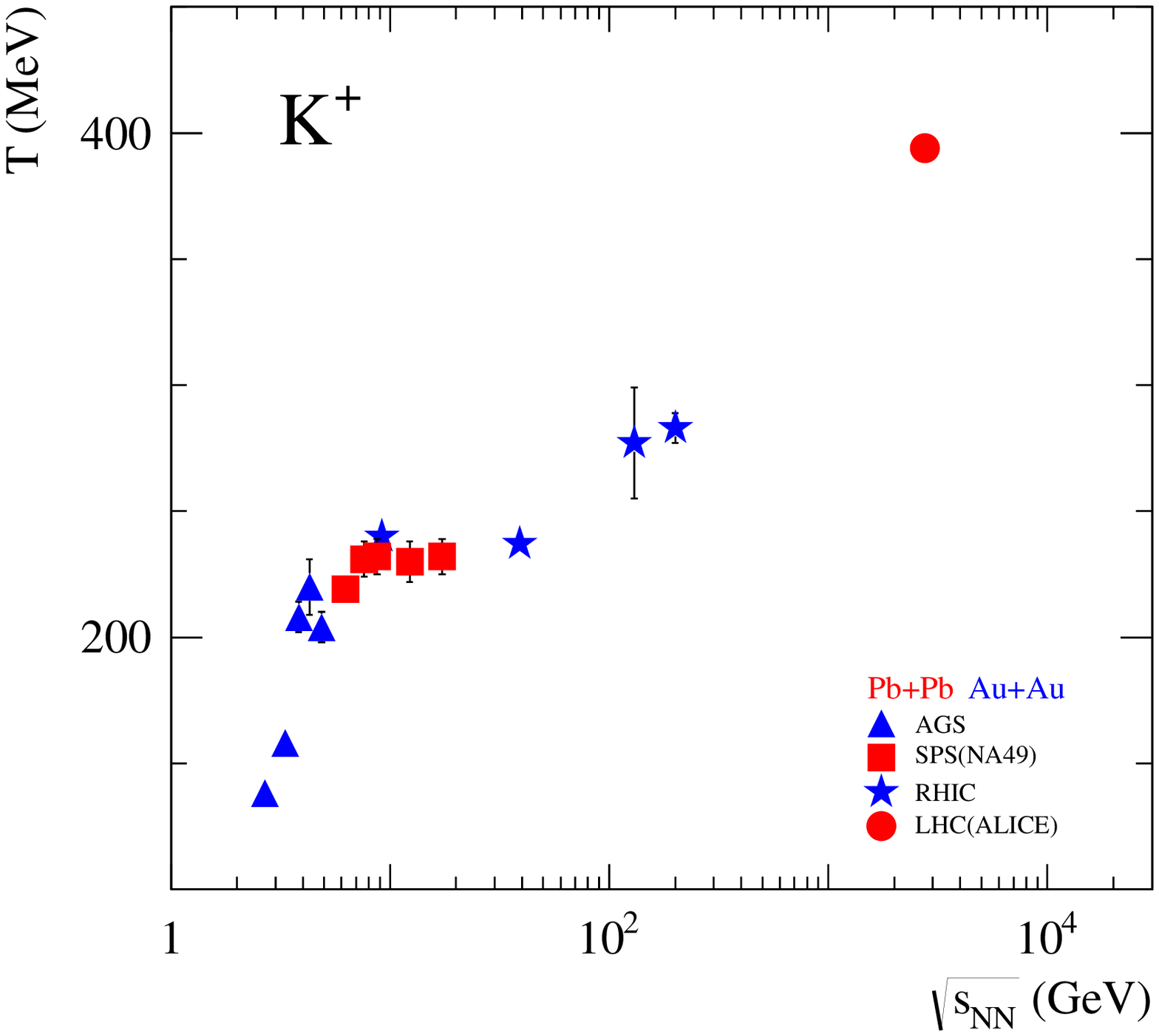}
\caption{\footnotesize Inverse slope parameters of kaon $m_T$ spectra.}
\label{fig2}
\end{figure}

\begin{wrapfigure}{r}{7.0cm}
\includegraphics[scale=.37]{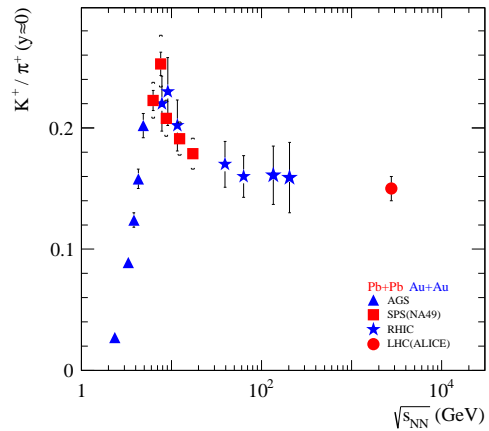}
\caption{\footnotesize Ratio of kaon to pion yield (near midrapidity).}
\label{fig3}
\end{wrapfigure}

Figure~\ref{fig2} shows inverse slope parameters of kaon transverse mass ($m_T - m_0$)
spectra. The LHC points and the RHIC BES points confirm the step structure expected for the onset of deconfinement.
The $K^{+}/\pi^{+}$ yield (near midrapidity) is presented in Fig.~\ref{fig3}.

As seen, RHIC results confirm NA49 measurements at the onset of deconfinement. Moreover, LHC (ALICE) data demonstrate that the energy dependence of hadron production properties shows rapid changes only at low SPS energies, and a smooth evolution is observed between the top SPS (17.3~GeV) and the current LHC (2.76~TeV) energies. All three structures confirm that results agree with the interpretation of the NA49 structures as due to onset of deconfinement. Above the onset energy only a smooth change of QGP properties with increasing energy is expected.

\section{New NA49 results on fluctuations}
Fluctuations and correlations may serve as a signature of the onset of deconfinement. Close to the phase transition the equation of state changes rapidly which can impact the energy dependence of fluctuations. Moreover,
fluctuations and correlations can help to locate the critical point of strongly interacting matter. This is in analogy to critical opalescence, where we expect enlarged fluctuations close to the critical point. For strongly interacting
matter a maximum of fluctuations is expected when freeze-out happens near the critical point. Therefore the critical point should be searched above the onset of deconfinement energy, found by NA49 to be 30A GeV ($\sqrt{s_{NN}}\approx 7.6~{\rm GeV}$).

\subsection{Proton and pion intermittency signals}
It was suggested that the analog of critical opalescence may be detectable through intermittency analysis in $p_{T}$ space. Significant $\sigma{\rm -field}$ fluctuations are expected at the critical point (density fluctuations of zero mass $\sigma{\rm -particles}$ produced in abundance at the critical point)~\cite{Antoniou:2005am}. This critical point is the endpoint of a line of first order transitions associated with the partial restoration
of the chiral symmetry when the temperature T, for given baryochemical potential $\mu_{B}$, increases beyond a critical value $T_{c}$. $\sigma$~particles at $T<T_{c}$ may reach the two-pion threshold ($2m_{\pi}$) and then decay into two pions, therefore density fluctuations of di-pions with $m_{\pi^{+}\pi^{-}}$ close to the two pion mass incorporate $\sigma {\rm -field}$ fluctuations at the critical point. Local density fluctuations are expected both in configuration and momentum space. In a finite-density medium there is a mixing between the chiral condensate
and the baryon density. Thus the critical fluctuations of the $\sigma{\rm -field}$ induce fluctuations of the baryon density~\cite{Fukushima:2010bq} as well. Furthermore, as pointed out in~\cite{Hatta:2003wn} the critical fluctuations
of the chiral condensate are also directly transferred to the net proton density through the coupling of
the protons with the isospin zero $\sigma{\rm -field}$. Since protons are easier to detect than neutrons the perspective of detecting the QCD critical point through fluctuations of the net proton density is very promising.

\begin{figure}
\includegraphics[width=.75\textwidth]{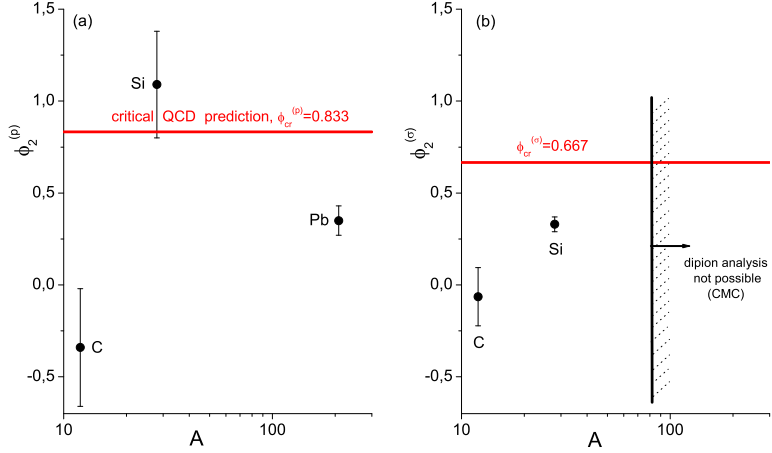}
\caption{\footnotesize (a): intermittency index $\phi_2$ for protons in 10\% most central C+C, Si+Si and Pb+Pb interactions at 158A GeV. (b): $\phi_2$ for low-mass $\pi^+\pi^-$ pairs in p+p, and 10\% most central C+C and Si+Si interactions at 158A GeV.}
\label{fig4}
\end{figure}

The NA49 experiment searched for an intermittency signal in transverse momentum space of reconstructed
di-pions ($\pi^{+}\pi^{-}$ pairs) with invariant mass just above $2m_{\pi}$~\cite{Anticic:2009pe} and protons~\cite{Anticic:2012pr}. The analysis was performed for p+p, C+C and Si+Si interactions (pion intermittency) and C+C, Si+Si and Pb+Pb (proton intermittency) both at 158A GeV. 
The second scaled factorial moments $F_{2}(M)$ (with M being the number of subdivisions in each transverse momentum space direction) of di-pion and proton densities in transverse momentum space were computed for real data and for artificially produced mixed events where only statistical fluctuations are present. The combinatorial background subtracted (by use of mixed events) moments $\Delta F_{2}$ in transverse momentum space are expected to follow a power-law behavior 
$\Delta F_{2}\sim (M^{2})^{\phi_{2}}$, with $\phi_{2}=2/3$ (di-pions), and $\phi_{2}=5/6$ (protons) for systems freezing-out at critical point~\cite{Antoniou:2005am}.

Figure~\ref{fig4} shows that the value of $\phi_2$ for $\Delta F_{2}$ in Si+Si collisions at the top SPS energy indicates fluctuations
approaching in size the prediction of critical QCD. The remaining departure may be due to freezing out at
a distance from the critical point.

\section{Summary}
The NA49 discovery of the energy threshold for deconfinement is now confirmed. The results from the RHIC Beam Energy Scan agree with NA49 measurements on the onset of deconfinement. LHC data confirm the interpretation 
of the structures observed at low SPS energies as due to onset of deconfinement.
For central A+A collisions fluctuations of pion and proton densities tend to a maximum in Si+Si collisions
at 158A GeV. Thus the critical point may be accessible at SPS energies.
This result is a strong motivation for future experiments and in fact, the NA49 efforts will be continued by the ion program of the NA61/SHINE experiment~\cite{Gazdzicki:2011fx}.

\end{document}